\documentclass[lettersize,journal]{IEEEtran}

\usepackage{pifont}
\usepackage{amsmath,amssymb,amsfonts}
\usepackage{graphicx}
\usepackage{booktabs}
\usepackage{multirow}
\usepackage{cite}
\usepackage{url}
\usepackage{xurl}
\usepackage[hidelinks]{hyperref}
\usepackage{algorithm}
\usepackage{algorithmic}
\usepackage{stfloats}
\usepackage{xcolor}
\usepackage{balance}
\usepackage{subfigure}
\begin{document}

\title{SimART: A Unified and Open Real-world Multimodal Simulation Platform for 6G Integrated Sensing and Communication}

\author{\IEEEauthorblockN{
        Kang Yan, 
        Yuqi Cao, 
        Jiaqi Li,
        Luping~Xiang, \emph{Senior~Member, IEEE},
		and~Kun~Yang, \emph{Fellow,~IEEE}}

    \thanks{Kang~Yan and Yuqi~Cao are with the School of Information and Communication Engineering, University of Electronic Science and Technology of China, Chengdu, 611731, China, email: kangyan@std.uestc.edu.cn; yuqicao@std.uestc.edu.cn. \textit{(Kang~Yan and Yuqi~Cao contributed equally to this work.)}}
	\thanks{Jiaqi~Li, Luping~Xiang and Kun~Yang are with the State Key Laboratory of Novel Software Technology, Nanjing University, Nanjing, 210008, China, Institute of Intelligent Networks and Communications (NINE), Collaborative Innovation Center of Novel Software Technology and Industrialization, and School of Intelligent Software and Engineering, Nanjing University (Suzhou Campus), Suzhou, 215163, China, email: 15520701520@163.com, luping.xiang@nju.edu.cn, kunyang@nju.edu.cn. \textit{(Corresponding author: Luping~Xiang.)}}
}

\maketitle

\begin{abstract}
Research on sixth-generation (6G) integrated sensing and communication (ISAC) increasingly depends on multimodal datasets. These datasets need to jointly characterize wireless propagation, onboard sensing, and platform mobility. Existing tools cover only part of these aspects. Robotics simulators model physics and perception but not site-specific channels, while ray tracing and link level tools lack vehicle dynamics and onboard sensors. Combining them manually leads to workflows that are fragile and hard to reproduce. Rather than introducing another standalone simulator, this article presents SimART. It integrates mature robotics, ray tracing, and wireless evaluation engines into a single reproducible pipeline. The key idea is a robot operating system (ROS) backbone that both synchronizes and organizes all multimodal streams. A shared clock, a common coordinate frame, and timestamped messages keep the streams aligned in time and space, and a single rosbag recording captures the full session into one reproducible file. This design decouples the sensing front end from the wireless back end, so that any ROS-compatible simulator can be plugged in while reusing the same back end across aerial, ground, indoor, and maritime ISAC settings. On top of this backbone, SimART contributes a scene construction pipeline that converts both OpenStreetMap extracts and user-defined layouts into spatially aligned visual and electromagnetic assets, and a channel knowledge map (CKM) generator that aggregates ray tracing and system level outputs into spatial priors for ISAC algorithms. A case study on vision  and position aided beam prediction demonstrates the utility of the platform. The code is publicly available at \underline{\url{https://github.com/guchuanv-alt/SimART}}.
\end{abstract}


\section{Introduction}

\IEEEPARstart{T}{he} development of sixth-generation (6G) wireless networks has made integrated sensing and communication (ISAC) a key research topic \cite{dang2020should}. ISAC scenarios span a wide range of mobile platforms and deployment environments. Aerial platforms, such as unmanned aerial vehicles, urban air mobility vehicles, and delivery drones, must sense their surroundings while maintaining high data rate links with ground infrastructure in three-dimensional (3D) urban environments \cite{jiang20236g}. Ground vehicles in connected and autonomous driving rely on joint sensing and vehicle-to-everything (V2X) communication. Indoor robots in smart factories and warehouses operate in dense multipath environments where perception and connectivity must be tightly coupled. Despite their diversity, research on these scenarios relies on scenario-specific datasets \cite{alkhateeb2023deepsense} that jointly capture visual sensing, inertial measurements, electromagnetic propagation, link level performance, and spatial channel characteristics with strict temporal alignment and geometric consistency.

Real-world measurement campaigns for ISAC are costly, subject to regulatory constraints, and difficult to reproduce \cite{khuwaja2018survey}, especially when multiple robotic platforms or large-scale environments are involved. As a result, simulation has become an important tool for algorithm development, dataset generation, and performance evaluation. However, existing platforms usually provide only a subset of the capabilities required by ISAC research. Robotics simulators, such as CARLA \cite{dosovitskiy2017carla}, AirSim \cite{shah2017airsim}, Gazebo \cite{koenig2004design}, and Isaac Sim \cite{makoviychuk2021isaac}, offer 3D physics, visual sensing, LiDAR simulation, robot operating system (ROS) integration, and customizable scene construction, but do not natively support wireless propagation modeling, link and system level wireless evaluation, or channel knowledge map (CKM) generation. Wireless oriented tools, such as Sionna RT/SYS \cite{sionna}, DeepMIMO \cite{alkhateeb2019deepmimo}, and Wireless InSite \cite{remcom2021wireless}, can generate ray tracing based channels or support wireless performance evaluation, but they generally lack vehicle dynamics, onboard perception sensors, and ROS-based robotic integration. Network simulators such as ns-3 \cite{riley2010ns} support link and system level network evaluation, but they operate at a higher abstraction level and do not directly model three dimensional scene geometry, platform motion, or multimodal sensing observations.

This fragmentation creates a practical barrier for generating scenario-specific ISAC datasets and evaluating ISAC algorithms across diverse environments. Researchers who need cross-modal datasets must often build custom interfaces among heterogeneous tools, run different simulators sequentially, export data in incompatible formats, and manually align timestamps and coordinate frames. Such workflows are fragile, difficult to reproduce, and hard to extend across different environments, robotic platforms, and communication and sensing tasks.

Rather than developing another standalone simulator, ISAC research requires an integration layer that connects mature robotics, ray tracing, and wireless evaluation tools in a consistent and reusable manner. Such a platform should maintain spatial and temporal consistency across sensing, mobility, wireless propagation, and link level performance evaluation. It should also support open and configurable scene construction, enabling researchers to study real urban areas, synthetic layouts, and user-defined environments within the same reproducible workflow.

This article presents SimART, a scenario-configurable multimodal simulation platform for 6G ISAC. The main contributions of this work are summarized as follows.

\begin{itemize}
\item A modular cross-simulator architecture is proposed to integrate physics and sensing, ray tracing based channel generation, link and system level wireless evaluation, and channel knowledge map (CKM) construction into a unified simulation pipeline. A ROS-compatible front-end interface, with a shared simulation clock, a common tf2 coordinate frame tree, and timestamped messages, enforces spatial and temporal consistency across modules and allows the same wireless and CKM back end to be reused with different robotics simulators, supporting aerial, ground, indoor, maritime ISAC scenarios, as well as user-defined ISAC scenarios.


\item A dual scene construction pipeline is introduced for both real-world and synthetic scenarios. The pipeline reconstructs real geographic areas from OpenStreetMap data with electromagnetic material assignment in Blender, while also supporting controlled user-defined layouts through custom mesh import and configurable scene interfaces.

\item A unified multimodal data collection mechanism is implemented to produce ISAC datasets that are synchronized, reproducible, and directly compatible with standard ROS tools. The complete session, including sensor streams, wireless channels, link-level KPIs, and CKM layers, can be captured into a single rosbag file for replay, inspection, and export to downstream learning tasks.
\end{itemize}



\begin{table*}[!t]
    \centering
    \caption{Comparison of representative simulation platforms.}
    \resizebox{7in}{!}{
    \begin{tabular}{l|c|c|c|c|c|c|c|c|c|c}
        \hline  
		\text {Platform} 
		& \text{3D Physics} 
		& \text{Vision} 
		& \text{LiDAR} 
		& \text{Ray Tracing} 
		& \text{Link/System} 
		& \text{CKM} 
		& \text{ROS} 
		& \text{Real Map} 
		& \text{Custom Scene} 
        & \text{Open Source} 
		\\
        \hline 

        \hline 
		\text{CARLA \cite{dosovitskiy2017carla}} 
		&{\ding{51}}&{\ding{51}}&{\ding{51}}
        &{}&{}&{}
        &{\ding{51}}&{\ding{51}}& {\ding{51}}&{\ding{51}}\\
        \hline 
        
		\text{AirSim \cite{shah2017airsim}}
		&{\ding{51}}&{\ding{51}}&{\ding{51}}
        &{}&{}&{}
        &{\ding{51}}&{\ding{51}}& {\ding{51}}&{\ding{51}}\\
        \hline
        
		\text{Gazebo \cite{koenig2004design}}
		&{\ding{51}}&{\ding{51}}&{\ding{51}}
        &{}&{}&{}
        &{\ding{51}}
        &{}
        &{\ding{51}}&{\ding{51}}\\
        \hline
        
		\text{Isaac Sim \cite{makoviychuk2021isaac}} 
		&{\ding{51}}&{\ding{51}}&{\ding{51}}
        &{}&{}&{}
        &{\ding{51}}
        &{}
        &{\ding{51}}&{\ding{51}}\\
        \hline

        \text{Sionna RT/SYS \cite{sionna}} 
		&{}&{}&{}
        &{\ding{51}}&{\ding{51}}&{\ding{51}}
        &{}
        &{\ding{51}}&{\ding{51}}&{\ding{51}}\\
        \hline
        
        \text{DeepMIMO \cite{alkhateeb2019deepmimo}} 
		&{}&{}&{}
        &{\ding{51}}&{\ding{51}}
        &{}&{}&{}&{}
        &{\ding{51}}\\
        \hline

        \text{WirelessInsite \cite{remcom2021wireless}} 
		&{}&{}&{}
        &{\ding{51}}
        &{}&{}&{}
        &{\ding{51}}&{\ding{51}}
        &{}\\
        \hline

        \text{ns-3 \cite{riley2010ns}} 
		&{}&{}&{}&{}
        &{\ding{51}}
        &{}&{}&{}
        &{\ding{51}}& {\ding{51}}\\
        \hline

        \textbf{SimART (ours)} 
		&{\ding{52}}&{\ding{52}}&{\ding{52}}
        &{\ding{52}}&{\ding{52}}&{\ding{52}}
        &{\ding{52}}
        &{\ding{52}}&{\ding{52}}& {\ding{52}}\\
        \hline
    \end{tabular}
	}
    \label{Contributions}
\end{table*}

\section{Platform Architecture}

\subsection{Overall Design}
\begin{figure*}
\centering
\includegraphics[width=6.5in]{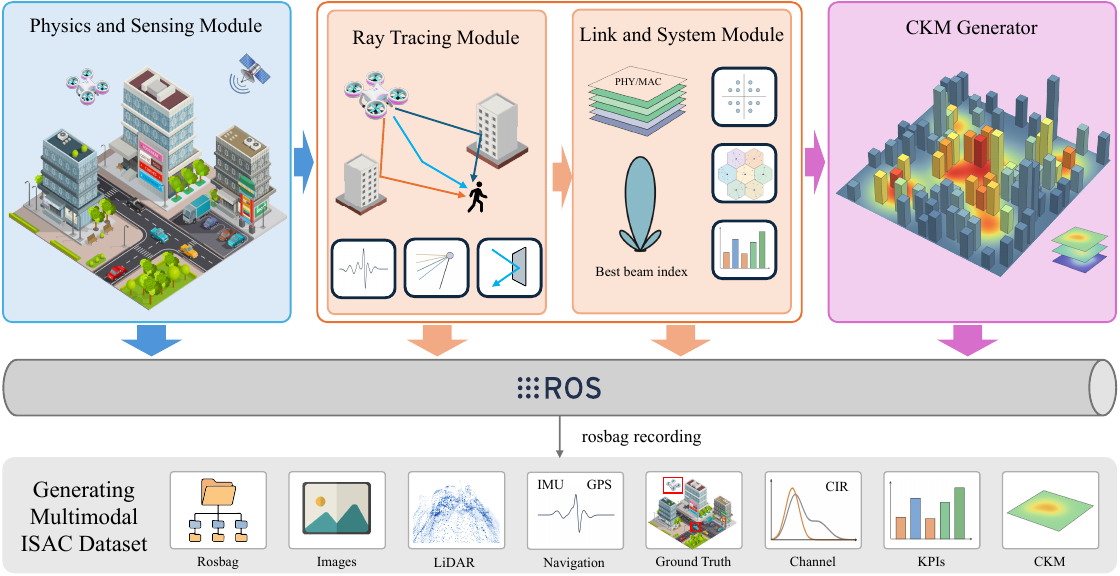}
\caption{Architecture of SimART and the resulting multimodal ISAC dataset.}
\label{SystemArchitecture}
\end{figure*}

Fig.~\ref{SystemArchitecture} illustrates the architecture of SimART. The platform consists of four functional modules: the Physics and Sensing module, the Ray Tracing module, the Link and System module, and the CKM generator. The first three modules are built on mature third-party simulation engines, while the CKM generator is implemented as a SimART module that aggregates ray tracing and system-level outputs. Each module is exposed through a unified ROS interface. All modules run as concurrent ROS nodes and exchange data through standardized messages, which decouples simulation engines from downstream modules and enables modular replacement. The detailed message definitions are presented in Section~\ref{SecRecordedData}.

\subsection{Physics and Sensing Module}
The physics and sensing module in SimART is defined by a ROS contract rather than by a specific simulator. Concretely, this module is required to publish platform pose on the tf2 tree and to expose the configured onboard sensors as standard ROS topics, including RGB images, depth images, semantic segmentation images, LiDAR point clouds, IMU, GPS, and ground truth poses. Any robotics simulator that satisfies this contract can serve as the front end. AirSim is used as the reference implementation in this work, while other mainstream candidates such as Gazebo, Isaac Sim, and CARLA can be integrated in the same way.

\subsection{Ray Tracing Module}
The ray tracing module built on Sionna RT loads a geometrically simplified version of the scene rendered by the physics and sensing module. This simplified scene preserves the same buildings, roads, and major structural elements, while removing fine architectural details such as window frames, rooftop fixtures, and vegetation clutter. This simplification is motivated by the different computational characteristics of visual rendering and electromagnetic ray tracing. Modern GPU rasterization is highly parallelizable, so visual rendering scales gracefully with mesh complexity. In contrast, ray tracing requires per-ray intersection tests, and its cost scales roughly linearly with the number of triangles in the scene. Therefore, retaining such fine details would significantly increase computational cost while contributing only minor diffuse scattering, with limited impact on the dominant specular and edge-diffraction paths. To preserve electromagnetic realism, the simplified mesh is annotated with the same surface material properties as the full mesh, including relative permittivity and conductivity.

Based on this mesh, the ray tracing module computes propagation paths between transmitters and receivers, including complex path amplitudes, propagation delays, angles of arrival, Doppler frequency shift, and interaction points. These per-path quantities are aggregated into channel impulse responses (CIRs) that serve as input to the link and system module.

The ray tracing module supports three operating modes:

\textbf{Online mode:} transmitter and receiver positions are streamed from the physics and sensing module via ROS, and propagation paths are computed and visualized as the simulation proceeds.

\textbf{Offline replay mode:} a previously recorded rosbag of trajectories is replayed to recompute the ray tracing data, enabling reproducible experiments and parameter sweeps over the same motion.

\textbf{Grid scan mode:} with the transmitter held fixed, paths are computed for a dense grid of virtual receivers across the region of interest. The resulting data serves as the raw input to the CKM generator.

\subsection{Link and System Module}
The link and system module is built on Sionna SYS. It takes the CIRs produced by the ray tracing module as input and simulates physical layer (PHY) and medium access control (MAC) layer behavior under realistic, site-specific channels. Configurable components include OFDM waveforms with adjustable numerology, multi-antenna transmit and receive chains, and digital beamforming codebooks.

The module outputs link level key performance indicators (KPIs) at each simulation step, including signal-to-interference-plus-noise ratio (SINR), block error rate (BLER), and achievable rate. For beam-management studies, it additionally reports the optimal beam index from a given codebook, where the optimal beam is defined as the one that maximizes the received SINR at each time instant. All outputs are timestamped and aligned with the corresponding CIRs and with the sensor data produced by the physics and sensing module, enabling direct association of link level performance with the visual and geometric context in which it is observed.

\subsection{CKM Generator}
The CKM generator addresses an emerging need in ISAC research: location specific channel priors that support sensing-assisted communication and communication-assisted sensing. In SimART, the generator discretizes the region of interest into a 2D ground-plane grid at a configurable, fixed receiver height. With the transmitter held at a fixed location, the generator iterates over all grid cells and, at each cell, places a virtual receiver at the cell center, drives the ray tracing module in grid scan mode to obtain the propagation paths, and then invokes the link and system module to evaluate link level performance under those paths.

The per cell outputs are reduced to a set of summary quantities spanning both physical channel statistics and link level metrics, including total path loss, root-mean-square (RMS) delay spread, angular spread, average SINR, achievable rate, and the optimal beam index. Each quantity forms one layer of the resulting multi-layer CKM. Although this exhaustive enumeration is computationally demanding, it produces deterministic, high-fidelity CKMs that depend only on the underlying ray tracing and link level models, providing a reliable reference for evaluating future learning-based CKM estimation methods. The current 2D implementation can be naturally extended to a 3D voxel grid by sweeping multiple receiver heights.

\subsection{ROS-Based Integration}
The ROS backbone is the key component that enables SimART to function as an integrated platform rather than a simple combination of independent tools. The consistency of the modules is enforced through three designs, without requiring any module to access the internal implementation of others.

\textbf{A shared simulation clock.} A dedicated clock node drives the entire platform in ROS simulated time mode, decoupling wall clock execution speed from simulated time progression. This is essential because the ray tracing module typically runs slower than real time. With a shared clock, sensor streams and channel outputs remain temporally aligned regardless of per-step rendering or ray tracing latency.

\textbf{A unified coordinate frame tree.} The world, transmitter, receiver, and sensor frames are all registered in a common tf2 tree rooted at a configurable world origin. Operations such as projecting LiDAR points into the camera frame or associating a channel sample with the current transceiver pose are handled through standard frame transformations, without custom coordinate-management code at the module level.

\textbf{Timestamped message passing.} Every message produced by a module carries a header timestamp derived from the shared clock, allowing ROS approximate time synchronization tools to align multimodal streams without additional user defined logic. A single rosbag recording can capture an entire session, including all sensor streams, channel and system outputs, and coordinate frames, into one file that can be replayed in downstream analysis or training pipelines.

\begin{figure}[!t]
\centering
\subfigure[]{
    \includegraphics[width=0.45\textwidth]{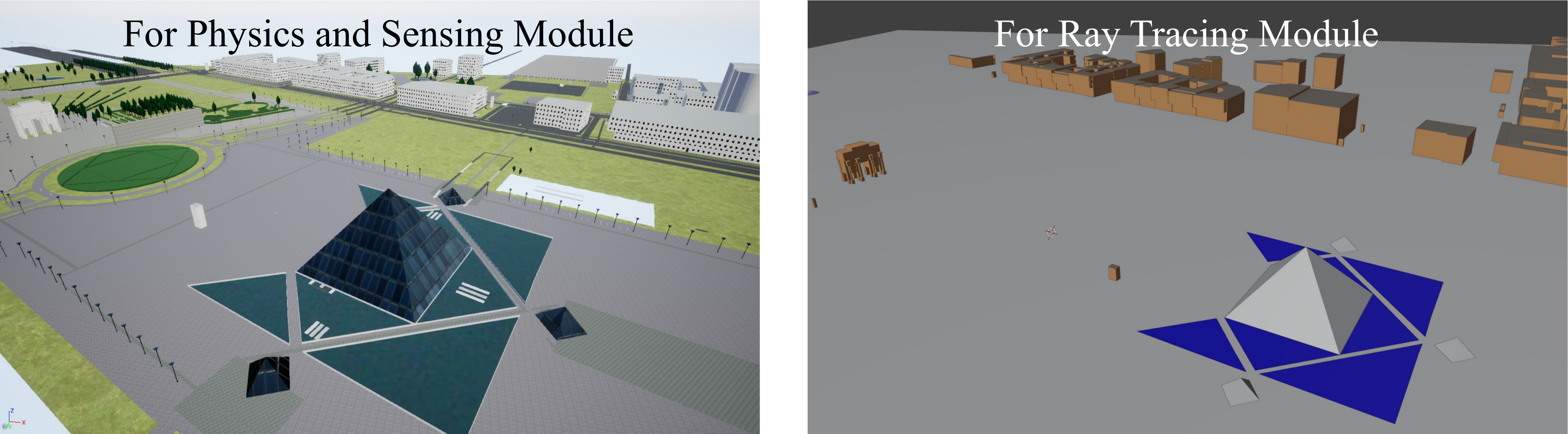}
}
\hfill  
\subfigure[]{
    \includegraphics[width=0.45\textwidth]{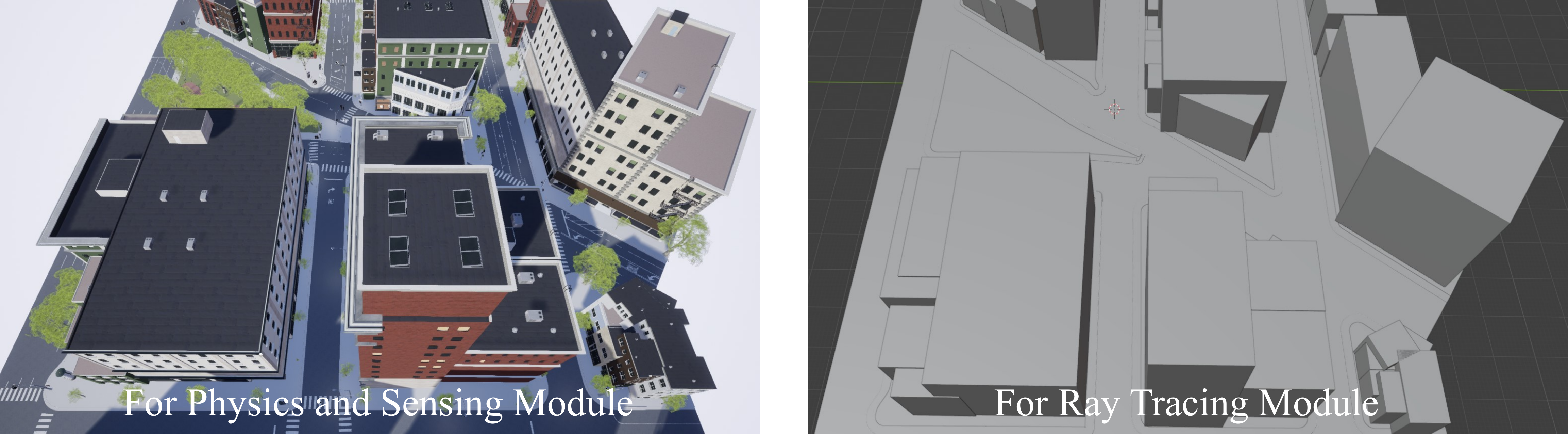}
}
\caption{Dual scene construction pipelines in SimART. The real-world pipeline reconstructs the Louvre area in Paris from OpenStreetMap data (a). The user-defined pipeline starts from a RoadRunner map \cite{mathworks_roadrunner_2023} and converts it into a propagation ready mesh through automatic simplification (b).}
\label{SceneConstruction}
\end{figure}

\section{Scene Construction and Map Adaptation}
Scene construction for ISAC simulation often involves two competing requirements. On one hand, studies conducted in real urban environments require high geographic fidelity, including realistic building footprints, road layouts, and vegetation that directly affect propagation. On the other hand, studies focused on algorithm design often require controlled environments in which individual factors, such as building density or street orientation, can be varied independently. To support both needs within a unified platform, SimART provides two complementary scene construction pipelines that share a common downstream interface, as shown in Fig.~\ref{SceneConstruction}. Regardless of how a scene is created, it is ultimately exported as two spatially aligned assets: a high fidelity scene asset for the physics and sensing module and a simplified mesh with material annotations for the ray tracing module. A shared world origin is embedded in both assets, so that transmitter and receiver poses are automatically registered in a common coordinate system.

\subsection{Real-World Map Adaptation Pipeline}
The real-world scene construction pipeline allows researchers to simulate transmitters and receivers in a specific city block, campus, or test field. It takes as input an OpenStreetMap (OSM) extract of the target area, which contains building footprints with height attributes, road networks, and selected land use polygons. Two parallel processing branches then convert this geographic description into the two assets required by SimART.

On the visual side, the OSM data are processed by OSM2World \cite{knerr2019osm2world}, which extrudes building footprints into 3D meshes, generates road surfaces with appropriate widths, and places default vegetation. The resulting geometry is imported into Unreal Engine as a static level, on top of which the physics and sensing module simulates platform motion dynamics and onboard sensing. On the wireless side, the same OSM extract is loaded into Blender and exported as a Mitsuba scene file for the ray tracing module. Material properties are assigned to OSM objects such as buildings, roads, and ground surfaces in Blender, following the material model adopted by Sionna RT.

Because the two branches share a common local coordinate frame, a position specified in this frame corresponds to the same physical location in both the physics and sensing module and the ray tracing module. This design eliminates per-scene manual alignment and avoids a common source of silent errors in multi-tool simulation pipelines.

\subsection{User Defined Scene Interface}
The user defined scene construction pipeline targets scenarios that cannot be readily extracted from OSM, such as controlled building grids for ablation studies, mixed indoor outdoor environments, or planned deployments that do not yet exist. Users construct the visual scene with any tool of their choice, such as RoadRunner or the Unreal Engine editor, and import the result as a static level for the physics and sensing module. To produce the corresponding ray tracing asset, SimART provides a Blender script that automatically converts the user built scene into a mesh compatible with the ray tracing module.

The script proceeds in three steps. First, fine geometric details, such as window frames, railings, and decorative structures, are removed based on object tags and size thresholds. Second, quadric edge-collapse decimation is applied to the remaining geometry while preserving sharp edges along building facades and ground planes, which dominate specular reflections. Third, each surface is assigned an electromagnetic material type from the Sionna RT material library.

The resulting mesh typically contains one to two orders of magnitude fewer triangles than the original visual scene, while preserving the large scale geometry that determines the dominant propagation paths. Because both assets are derived from a single user built source, they share a common local coordinate frame, preserving the spatial consistency of the SimART scene assets without additional user effort.

\section{Data Organization and Usage}

\subsection{Recorded Data}\label{SecRecordedData}
During a simulation session, each module publishes its outputs as ROS topics under a shared clock. A single rosbag record command captures the complete session into one file, thereby preserving the full multimodal stream of the session for replay, inspection, or export to other formats without rerunning any simulator. 
The sensing streams, including RGB images, depth images, semantic images, LiDAR, IMU, GPS, and positions, follow the standard ROS sensor messages and navigation messages conventions, making the recorded data directly compatible with established ROS tools for perception, SLAM, and robotics. The wireless streams, including per link channel impulse responses, link level key performance indicators, and selected beam indices, are defined through custom messages in a dedicated simart messages package. Since all streams share synchronized timestamps and a common tf2 frame tree, the association between a visual observation and its corresponding channel or beam state is well-defined by construction.

\begin{figure*}[!t]
	\centering
	
	\subfigure[]{
		\includegraphics[width=0.3\textwidth]{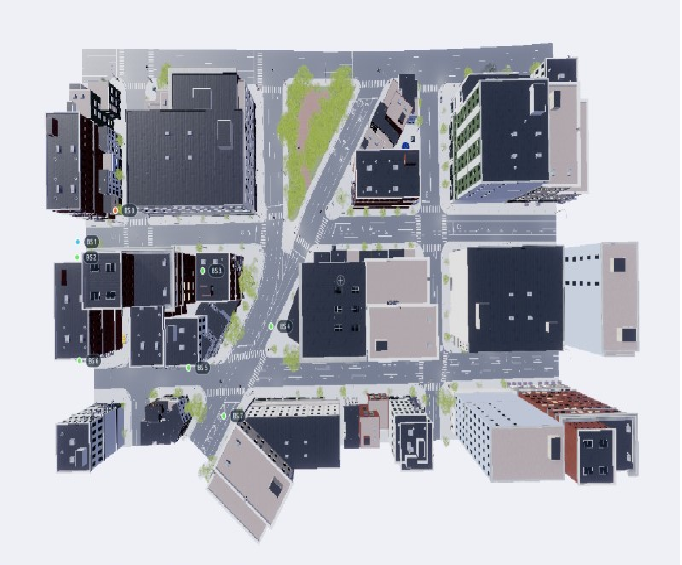}
        \label{data_overviewa}
	}
	\hfill
	\subfigure[]{
		\includegraphics[width=0.218\textwidth]{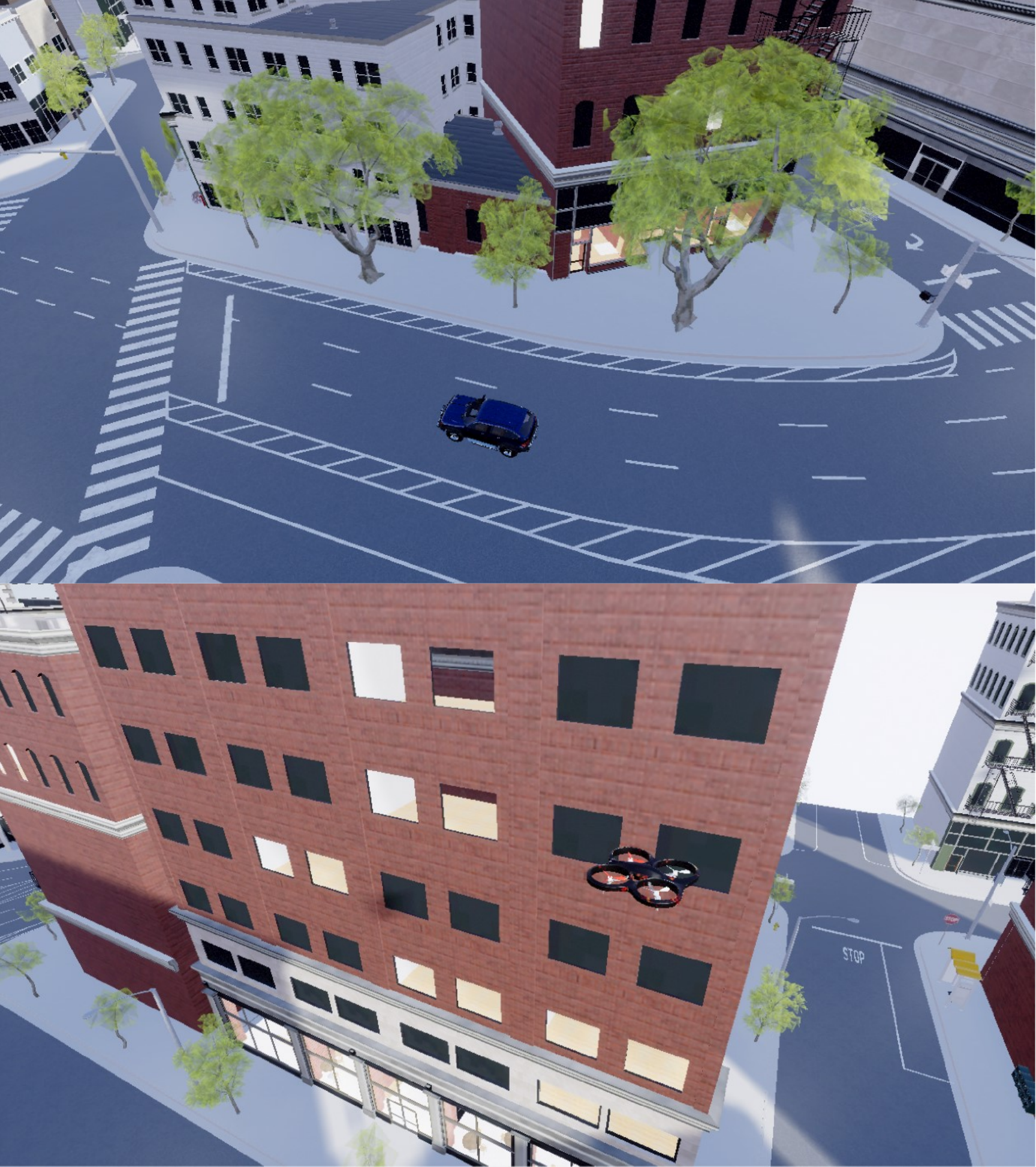}
        \label{data_overviewb}
	}
	\hfill
	\subfigure[]{
		\includegraphics[width=0.3\textwidth]{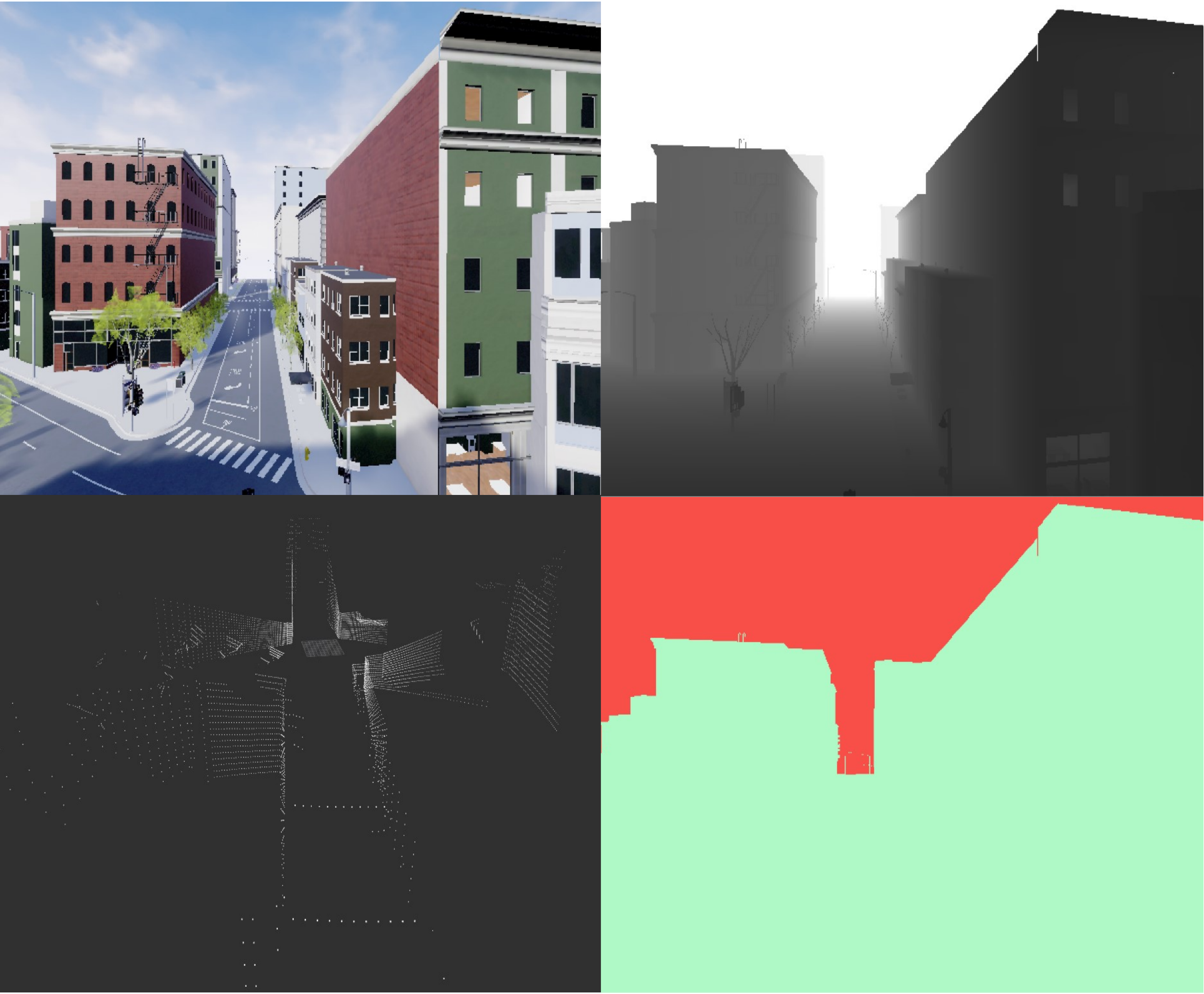}
        \label{data_overviewc}
	}
	
	\vspace{2mm}
	
	\subfigure[]{
		\includegraphics[width=0.3\textwidth]{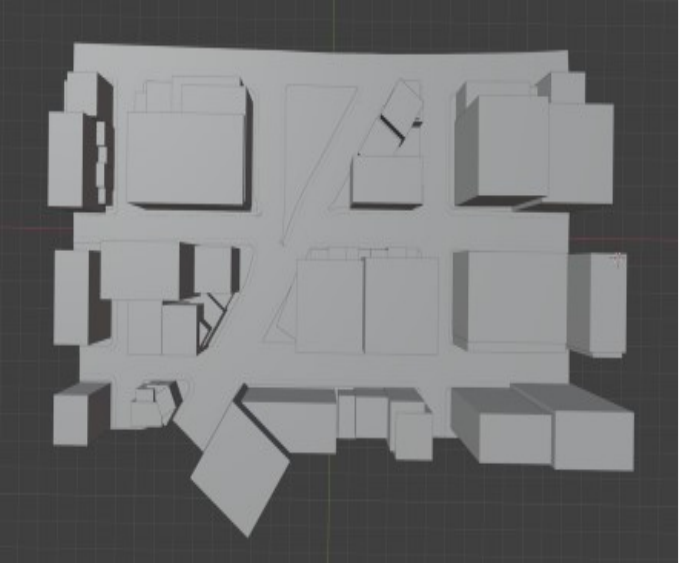}
        \label{data_overviewd}
	}
	\hfill
	\subfigure[]{
		\includegraphics[width=0.218\textwidth]{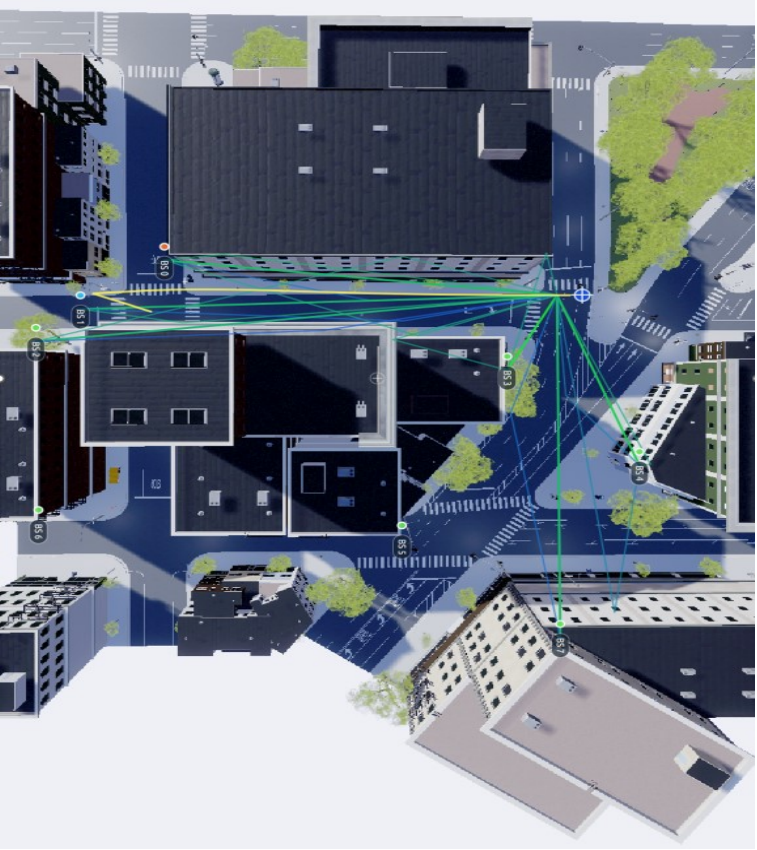}
        \label{data_overviewe}
	}
	\hfill
	\subfigure[]{
		\includegraphics[width=0.3\textwidth]{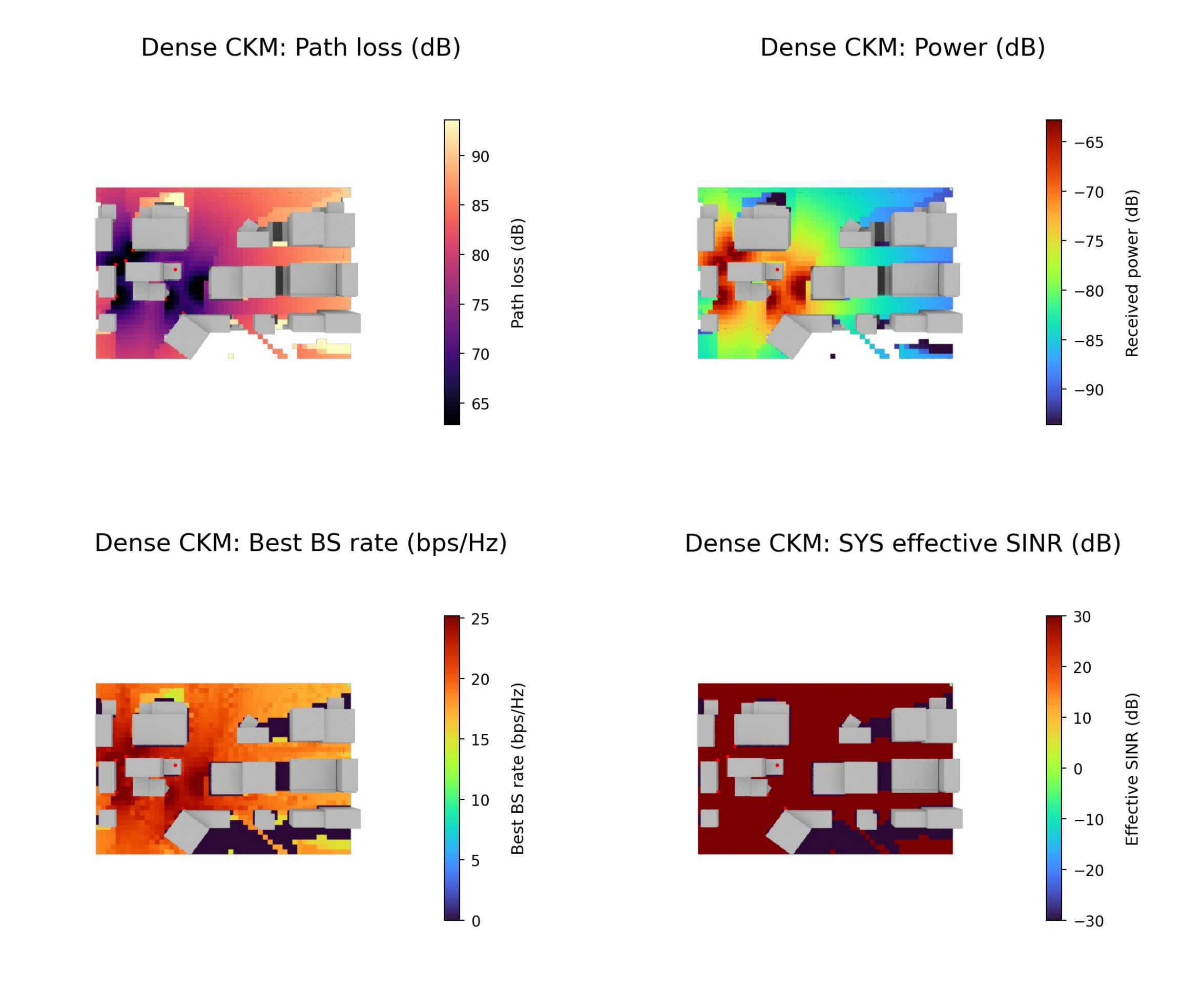}
        \label{data_overviewf}
	}
	
	\caption{Multimodal data produced by SimART in a representative session. (a) Visual scene loaded into AirSim for the physics and sensing module. (b) RGB view captured by the base-station camera, observing a car and a UAV. (c) Onboard UAV sensor outputs: RGB image, depth map, LiDAR point cloud, and semantic segmentation. (d) Simplified scene mesh used by the ray tracing module. (e) Ray tracing visualization of propagation paths between the base station and a virtual receiver. (f) CKM layers generated by the CKM generator: path loss, received power, best-BS achievable rate, and effective SINR.}
	\label{data_overview}
\end{figure*}

Fig.~\ref{data_overview} provides a visual summary of the multimodal data captured in a representative SimART session. The top row shows the outputs of the physics and sensing module: Fig.~\ref{data_overviewa} the visual scene loaded into AirSim, Fig.~\ref{data_overviewb} the RGB view captured by the base station mounted camera observing a car and a UAV, and Fig.~\ref{data_overviewc} the onboard sensor streams from the UAV, including RGB image, depth map, LiDAR point cloud, and semantic segmentation. The bottom row shows the wireless side: Fig.~\ref{data_overviewd} the simplified scene mesh used by the ray tracing module, Fig.~\ref{data_overviewe} a ray tracing visualization of the propagation paths between the base station and a virtual receiver, and Fig.~\ref{data_overviewf} CKM layers generated by the CKM generator, including path loss, received power, best BS achievable rate, and link level effective SINR. All quantities share the same simulation clock and coordinate frame, illustrating the spatial and temporal consistency enforced by the ROS backbone.

\subsection{Minimal Usage Example}
A typical SimART session is fully specified by a single configuration file, which defines the scene, sensor sampling rates, transmitter and receiver settings for Sionna RT, physical-layer numerology for Sionna SYS, and the voxel grid for CKM generation when applicable.

We illustrate a typical workflow using offline replay mode, in which the simulation is driven by a previously recorded trajectory rosbag containing the timestamped poses of all transmitters and receivers. Launching a session requires two commands: one to start the trajectory rosbag player, and one to launch the SimART simulation nodes. The simulation consumes the streamed poses, runs ray tracing and link level evaluation in lockstep with the shared clock, and publishes synchronized sensor, channel, and system outputs. When the trajectory rosbag finishes playing, the session can be configured to terminate automatically.

The complete session, including sensor streams, channel and system outputs, and coordinate frame transformations, can be captured into a single session rosbag via standard rosbag record. This recorded bag can be replayed by any ROS program for downstream analysis, training, or evaluation. Standard tools such as rqt bag and rviz additionally allow researchers to inspect images, point clouds, trajectories, and channel statistics from the session in a synchronized manner.

\section{Case Study: Multimodal Beam Prediction}
To demonstrate how SimART supports end-to-end ISAC research, we present a compact case study on vision and position aided beam prediction for a low-altitude communication link. In this setting, a ground-side base station serves a UAV flying through an urban block, while an RGB camera collocated with the base station captures visual observations of the airspace. The UAV location is detected from RGB images using a YOLOv8 network \cite{sohan2024review} and is combined with synchronized GPS information to predict the optimal transmit beam from a fixed codebook. The purpose of this case study is not to develop a new prediction algorithm, but to show that SimART can generate a self-consistent multimodal dataset and that a standard baseline trained on this dataset exhibits the expected behavior under different environmental conditions.

\subsection{Data Collection}
In this case study, we evaluate SimART using the user defined scene construction pipeline. The scene is built in RoadRunner with designed road and building layouts to provide a controllable virtual environment, and is then converted into paired AirSim and Sionna RT assets that share a common local coordinate frame. As a result, visual sensing, UAV motion, and wireless propagation are generated from spatially aligned representations of the same environment.

To evaluate the robustness of the generated multimodal data under different environmental conditions, data are collected under three representative scene settings: sunny daytime, rainy daytime, and nighttime. These settings share the same geometric layout and UAV trajectory design, but differ in both visual appearance and ray-tracing material configuration. Specifically, the AirSim rendering conditions are adjusted to reflect different illumination and weather effects, while the Sionna RT scene uses condition dependent material settings where applicable, especially to reflect dry and wet surface conditions.
Therefore, the resulting dataset captures not only visual domain shifts, but also corresponding changes in the wireless propagation characteristics.

A base station is mounted on a building rooftop at 27 m above ground level, equipped with an $8\times 8$ uniform planar array operating at 3.5 GHz. It serves a single-stream UAV terminal through a 64-beam DFT codebook. The UAV flies at altitudes between 1 m and 20 m along predefined trajectories that traverse line-of-sight regions with respect to the base station. An RGB camera collocated with the base station records synchronized visual observations of the airspace. In each RGB frame, the UAV is detected using a YOLOv8 network, which provides the image-domain UAV location. Meanwhile, the UAV GPS position is recorded and synchronized with the visual observation and ray tracing based beam label. The detected visual location and GPS position are jointly used as multimodal features for beam prediction.

\subsection{Performance}

Table~\ref{BeamPredResults} summarizes the vision and position aided beam prediction performance under different environmental conditions. The YOLOv8-based visual front end is used to detect the UAV location in each RGB frame. The resulting image-domain location is then concatenated with synchronized GPS position information and fed into the beam prediction network. The beam prediction network is trained using the combined data from sunny, rainy, and nighttime scenes, and is evaluated separately on each scene setting.

The results show that a standard multimodal baseline trained on the SimART dataset achieves consistently high beam prediction accuracy across all three conditions. In the sunny scene, the model achieves a top-1 accuracy of $97.86\%$, a top-3 accuracy of $99.83\%$, and a top-5 accuracy of $100\%$. In the rainy scene, the top-1 accuracy decreases moderately to $95.39\%$, while the top-3 and top-5 accuracies remain very high at $99.84\%$ and $100\%$, respectively. In the nighttime scene, the model achieves a top-1 accuracy of $97.92\%$, a top-3 accuracy of $99.94\%$, and a top-5 accuracy of $100\%$.

The high accuracy is expected because the predictor uses both image-domain UAV detection and GPS position information, which provides strong spatial cues for beam selection.
The slightly lower top-1 accuracy in the rainy scene is reasonable because this setting introduces a more challenging domain shift. On the visual side, rain changes image appearance and may affect the reliability of UAV detection. On the wireless side, the corresponding RT material configuration also slightly perturbs the ray-traced propagation characteristics, which may shift the optimal beam in a small fraction of frames. Nevertheless, the high top-3 and top-5 accuracies across all conditions indicate that the correct beam remains highly ranked even when both the visual observations and propagation environment vary, with the top-5 candidates always containing the correct beam.

\begin{table}[!t]
\centering
\caption{Vision and position aided beam prediction performance under different environmental conditions.}
\label{BeamPredResults}
\begin{tabular}{l|c|c|c}
\hline
\textbf{Scene Setting} & \textbf{Top-1} & \textbf{Top-3} & \textbf{Top-5} \\
\hline \hline
Sunny daytime & 97.86\% & 99.83\% & 100.00\% \\
\hline
Rainy daytime & 95.39\% & 99.84\% & 100.00\% \\
\hline
Nighttime & 97.92\% & 99.94\% & 100.00\% \\
\hline
\end{tabular}
\end{table}

\section{Conclusion}
This article presented SimART, a multimodal simulation platform for ISAC. Instead of introducing another standalone simulator, SimART is designed as an open integration layer in which photorealistic physics and sensing, ray tracing based wireless propagation, link and system level evaluation, and channel knowledge map generation are coupled only through ROS messages. As a result, the physics and sensing module is freely interchangeable among ROS-compatible simulators such as AirSim, Gazebo, and Isaac Sim, and SimART therefore inherits their combined scenario coverage, spanning aerial, ground, indoor, and maritime ISAC settings. AirSim is used as the reference implementation throughout this work for validation purposes. Spatial and temporal consistency across modalities is ensured by construction through a shared simulation clock, a common coordinate frame tree, and timestamped message passing.

\footnotesize
\balance
\bibliography{reference}

\end{document}